# Digital Nonlinearity Compensation in High-Capacity Optical Fibre Communication Systems: Performance and Optimisation

Tianhua Xu
*Connected Systems Group,
School of Engineering
University of Warwick
Coventry, UK*
tianhua.xu@warwick.ac.uk

Nikita A. Shevchenko
*Optical Networks Group,
Department of Electronic &
Electrical Engineering
University College London
London, UK*
mykyta.shevchenko.13@ucl.ac.uk

Boris Karanov
*Optical Networks Group,
Department of Electronic &
Electrical Engineering
University College London
London, UK*
boris.karanov.16@ucl.ac.uk

Gabriele Liga
*Optical Networks Group,
Department of Electronic &
Electrical Engineering
University College London
London, UK*
gabriele.liga.11@ucl.ac.uk

Domaniç Lavery
*Optical Networks Group, Department of
Electronic & Electrical Engineering
University College London
London, UK*
d.lavery@ucl.ac.uk

Robert I. Killey
*Optical Networks Group, Department of
Electronic & Electrical Engineering
University College London
London, UK*
r.killey@ucl.ac.uk

Polina Bayvel
*Optical Networks Group, Department of
Electronic & Electrical Engineering
University College London
London, UK*
p.bayvel@ucl.ac.uk

(Invited)

*Abstract*—Meeting the ever-growing information rate demands has become of utmost importance for optical communication systems. However, it has proven to be a challenging task due to the presence of Kerr effects, which have largely been regarded as a major bottleneck for enhancing the achievable information rates in modern optical communications. In this work, the optimisation and performance of digital nonlinearity compensation are discussed for maximising the achievable information rates in spectrally-efficient optical fibre communication systems. It is found that, for any given target information rate, there exists a trade-off between modulation format and compensated bandwidth to reduce the computational complexity requirement of digital nonlinearity compensation.

*Keywords—optical communication, achievable information rate, digital nonlinearity compensation, modulation format, digital back-propagation*

## I. INTRODUCTION

Optical networks form an integral part of the world-wide communication infrastructure and nowadays over 95% of all digital data traffic is carried over optical fibres. Achievable information rates (AIRs), which are natural figures of merit in coded communication systems for demonstrating the net data rates achieved [1-4], have increased greatly over the past few decades with the development of multiplexing techniques, improved optical fibres and amplifiers, advanced modulation formats, detection schemes and digital signal processing (DSP) [5-14]. These technologies together facilitated the revolution of the communication systems and the rapid growth of the Internet. Currently, optical fibre communications are challenged to meet the massive surge of information rate demands. However, the presence of Kerr effects in fibre medium is widely believed to impose an ultimate limit on further enhancement of the achievable information rates in optical communication systems. The signal distortions as an effect of the fibre nonlinearity are more significant in the systems that utilise larger transmission bandwidths, closer channel spacing as well as higher-order modulation formats [15-20].

Significant research efforts have been concentrated on the mitigation of the nonlinearity-induced degradations in optical transmission. A number of nonlinear compensation (NLC) techniques have been investigated, such as digital back-propagation (DBP), nonlinear pre-distortion, Volterra equalisation, optical phase conjugation, nonlinear Fourier transform and twin-wave phase conjugation, etc. [21-28]. Single-channel DBP, i.e. compensating only for intra-channel nonlinearities (self-phase modulation), has been suggested as a low-complexity compensation scheme which may realise potential cost savings in newly deployed systems [29-31]. Nevertheless, for a substantial increase in the achievable information rates, multi-channel DBP (MC-DBP) has been widely considered as a promising approach as it compensates for both intra-channel and inter-channel fibre nonlinearities in dense wavelength division multiplexed (WDM) optical communication systems [32-39]. In this paper, the performance and the optimisation of MC-DBP is discussed from the perspectives of signal-to-noise ratio (SNR) and AIR enhancement in the compensated fibre-optic communication systems, where different high-cardinality modulation formats such as dual-polarisation quadrature phase shift keying (DP-QPSK), dual-polarisation 16-level quadrature amplitude modulation (DP-16QAM), DP-64QAM and DP-256QAM, are applied. Both numerical simulations and analytical modelling have been carried out in a representative communication system consisting of nine 32-Gbaud Nyquist-spaced WDM channels transmitted over a standard single-mode fibre (SSMF) link.

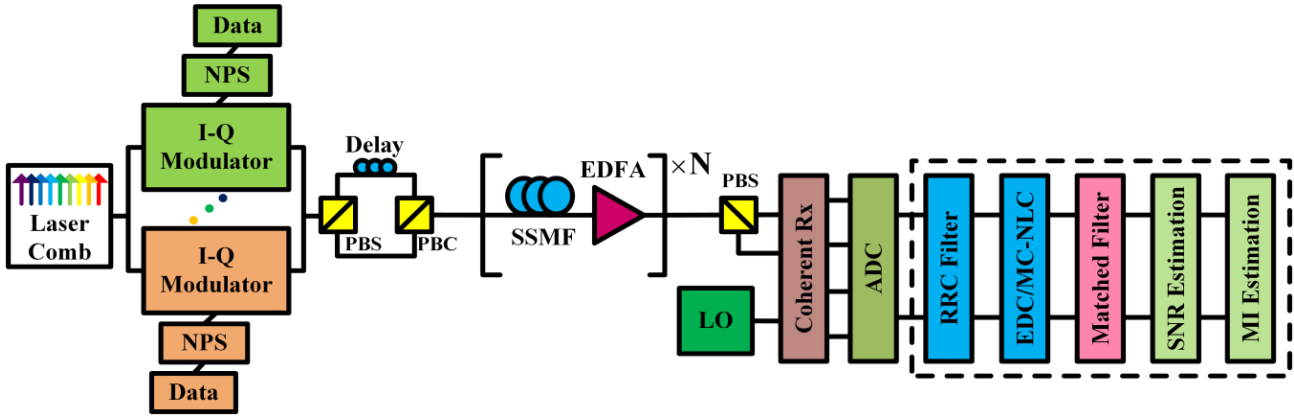

Fig. 1. Schematic of Nyquist-spaced optical communication system using multi-channel digital nonlinearity compensation (MC-NLC).

In the case where full-field DBP is applied to compensate for all nonlinear signal-signal interactions our investigation shows that the remaining distortions due to signal-noise beating become modulation format independent. On the other hand, the presence of signal-signal nonlinearity, when only an electronic dispersion compensation (EDC) or a partial-bandwidth DBP is utilised, exhibits a considerable dependence on the modulation format applied.

Regarding the backpropagation optimisation, when AIRs are considered the minimum required number of steps per span (MRNSPS) parameter in the DBP algorithm demonstrates a strong dependence on the modulation format for different back-propagated bandwidths. This is in contrast to the obtained MRNSPS with respect to SNR optimisation which does not show any modulation format dependence in the multi-channel DBP schemes under investigation.

Our study indicates that for any given target information rate the complexity requirement of the MC-DBP compensation can potentially reduce, benefitting from the existing trade-off between modulation format and back-propagated bandwidth.

## II. TRANSMISSION SETUP

The numerical setup of the 9-channel 32-Gbaud Nyquist-spaced superchannel optical transmission system is illustrated in Fig. 1. The applied modulation formats are DP-QPSK, DP-16QAM, DP-64QAM and DP-256QAM. The phase-locked optical carriers in the transmitter are generated using a 9-line 32-GHz spaced laser comb, and are de-multiplexed before I-Q optical modulators. The data in each channel are independent and random, and the symbols are further de-correlated with a delay of half the sequence length in the two orthogonal polarisations. The Nyquist pulse shaping (NPS) is realised using a root-raised cosine (RRC) filter with a roll-off of 0.1%. The SSMF span is simulated using the Manakov equation solved by the split-step Fourier method with a logarithmic distribution of the step size [40,41]. The erbium-doped optical fibre amplifier (EDFA) is applied to compensate for the loss in each fibre span. At the receiver end, the optical signals are mixed with the local oscillator (LO) laser to implement an ideal coherent detection.

TABLE I. TRANSMISSION SYSTEM PARAMETERS

| Parameter | Value |
|---|---|
| Symbol rate | 32 Gbaud |
| Channel spacing | 32 GHz |
| Central wavelengths (both Tx and LO) | 1550 nm |
| Number of channels | 9 |
| Roll-off | 0.1 % |
| Attenuation coefficient | 0.2 dB/km |
| Chromatic dispersion coefficient | 17 ps/nm/km |
| Nonlinear coefficient | 1.2 /W/km |
| Span length | 80 km |
| Number of spans | 25 |
| SSMF steps per span (logarithmic step-size) | 800 |
| EDFA noise figure | 4.5 dB |

In the DSP part, EDC is realised using frequency-domain equalisation [42,43], and the MC-DBP is performed using an reverse split-step Fourier solution of the Manakov equation [21,33,35]. An ideal RRC filter is further employed to change the back-propagated bandwidth in the MC-DBP module. The considered back-propagated bandwidths range from 32-GHz (1-channel NLC) to 288-GHz (full-field NLC). The matched filter then selects the central channel and removes the crosstalk from neighbouring channels. The SNR of the central channel is evaluated over $2^{18}$ symbols, and the mutual information (MI) is calculated from the obtained SNR, as discussed in [2,31,38]. A digital resolution is 2 sample/symbol/channel to guarantee the accuracy of numerical simulations. The phase noise from the transmitter and LO lasers, the frequency offset between them, as well as the polarisation mode dispersion (PMD) in the optical fibres are neglected. Table 1 shows system parameters in detail.

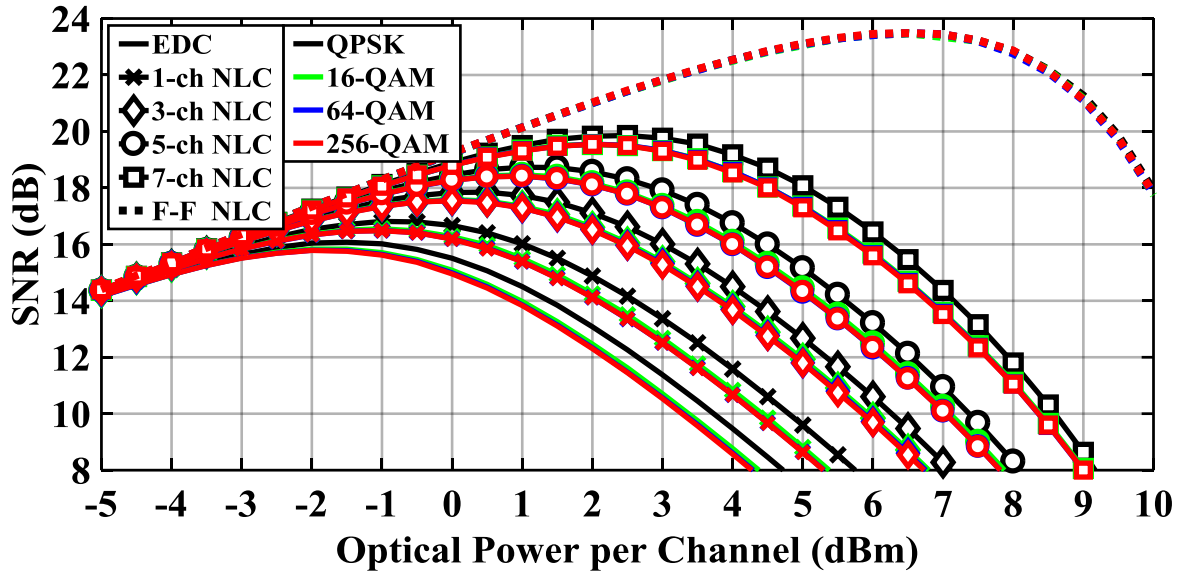

Fig. 2. Signal-to-noise ratio (SNR) versus optical launch power for systems with linear and multi-channel nonlinearity compensation.

### III. OPTIMISATION OF MULTI-CHANNEL DIGITAL BACK-PROPAGATION

Traditionally, the optimisation of MC-DBP algorithm was performed in terms of $Q^2$ factors and SNR performance [32-34]. However, in coded transmission systems, AIRs are more useful indicators which give a measure of the net data rates that can be achieved [1-4]. In this work, the multi-channel DBP algorithm is optimised in terms of both SNR and AIR values, where the minimum required number of steps per span for different back-propagated bandwidths and modulation formats are investigated in the considered 9-channel 32-Gbaud Nyquist-spaced optical communication system. Table 2 summarises the minimum required number of steps per span under both AIR and SNR optimisation criteria. The obtained results indicate that for a given back-propagated bandwidth in the MC-DBP scheme, the minimum required number of steps per span that achieves the highest possible AIR is dependent on the modulation format applied. Therefore, when optimisation is carried out for maximising the information rates, the choice of modulation format has an integral role. On the other hand, if MC-DBP is optimised with respect to the SNR, there is no modulation format dependence. As a consequence, the implications of the AIR optimisation are significant reductions of MRNSPS with decreasing order of modulation cardinality as compared to the SNR-optimised cases. For instance, full-field DBP maximises the AIR of the DP-QPSK systems when it is performed with 100 steps/span, while the SNR is maximised at 500 steps/span.

The comparison between the AIR- and SNR-optimised MRNSPS suggests that the conventional SNR optimisation often overestimates the system requirements. Therefore, in practice the complexity of applying MC-DBP may be considerably reduced if the AIRs instead of the SNRs are maximised for given modulation format.

TABLE II. MINIMUM REQUIRED NUMBER OF STEPS PER SPAN

| MRNSPS in terms of AIRs | | | | | |
|---|---|---|---|---|---|
| Formats | 32-GHz | 96-GHz | 160-GHz | 224-GHz | 288-GHz |
| DP-QPSK | 1 | 2 | 5 | 25 | 100 |
| DP-16QAM | 1 | 10 | 25 | 75 | 200 |
| DP-64QAM | 5 | 25 | 75 | 150 | 250 |
| DP-256QAM | 5 | 25 | 75 | 150 | 500 |
| MRNSPS in terms of SNRs | | | | | |
| DP-QPSK | 5 | 25 | 75 | 150 | 500 |
| DP-16QAM | 5 | 25 | 75 | 150 | 500 |
| DP-64QAM | 5 | 25 | 75 | 150 | 500 |
| DP-256QAM | 5 | 25 | 75 | 150 | 500 |

### IV. RESULTS AND DISCUSSIONS

In this section, the performance of MC-DBP is studied with respect to both SNR and AIR. The transmission link is 2000 km (25 span×80 km) SSMF. The nonlinear coefficient and the number of steps per span in the MC-DBP module are always the same as those in the transmission fibre to guarantee an optimal operation of MC-DBP.

Simulation results of SNR versus optical signal power per channel at different back-propagated bandwidths and modulation formats are shown in Fig. 2. The results show that in the cases of EDC and up to 7-channel NLC, the DP-

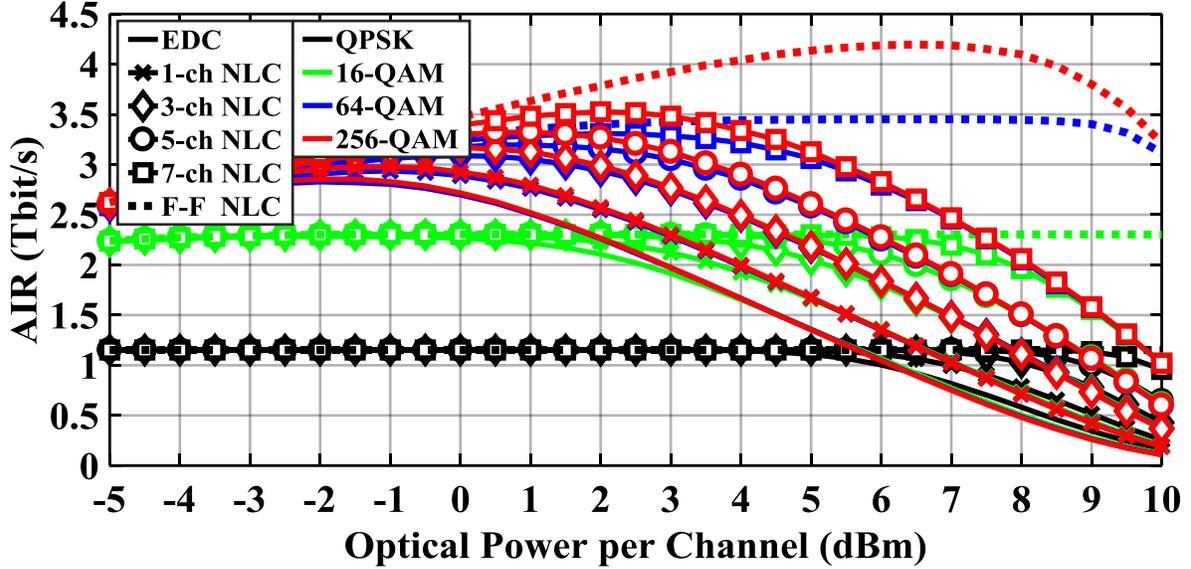

Fig. 3. AIRs versus input optical power per channel for systems with linear and multi-channel nonlinearity compensation.

16QAM, DP-64QAM and DP-256QAM systems have identical SNR performance, while the DP-QPSK system outperforms the other three modulation formats. This indicates that the nonlinear distortions in the cases of EDC and partial-bandwidth NLC, which are mainly from signal-signal interactions, depend on modulation format, and signal-signal interactions in the DP-QPSK system are smaller than in the systems using other higher-level modulation formats. This conclusion is consistent with earlier reports [44-46]. However, in the case of full-field NLC, all systems show similar SNR behaviour independent of the modulation formats applied. This demonstrates that the remaining nonlinear distortions for full-field NLC, which are mainly signal-noise interactions, do not depend on modulation formats.

Fig. 3 shows the simulated AIRs as a function of optical signal power per channel for the investigated transmission system at different back-propagated bandwidths, where different modulation formats are applied. It is found that in terms of AIRs the highest gain for full-field NLC is achieved at the highest-order modulation format (DP-256QAM), and is 1.34 Tbit/s (from 2.86 Tbit/s at -2 dBm in EDC to 4.20 Tbit/s at 6.5 dBm in full-field NLC). Furthermore, it can be observed in Fig. 3 that the AIR of the DP-256QAM system using 7-channel NLC is higher than that in the DP-64QAM system using full-field NLC, if the signal power is less than 3.5 dBm. For any examined power, the AIR of the DP-256QAM system using 7-channel NLC exceeds the AIRs of the full-field NLC cases in the DP-16QAM and DP-QPSK systems. Therefore, to achieve a given target AIR, a compromise could be made between the modulation format selection and the back-propagated bandwidth. The complexity of signal processing schemes may be traded for modulation generation complexity. In general, the effects of such a trade-off will depend on the particular transmission distance.

## V. CONCLUSIONS

In this paper, both the optimisation and performance of digital MC-NLC was studied from the perspectives of SNR and AIR in the long-haul optical fibre communication systems, when different modulation formats are applied. Numerical simulations were carried out in a 9-channel 32-Gbaud Nyquist-spaced optical communication system based on the SSMF transmission. Our results show that nonlinear distortions in the case of full-field NLC, which arise from signal-noise interactions, exhibit no modulation format dependence. On the other hand, nonlinear distortions in the system using EDC and partial-bandwidth NLC, which mainly arises from signal-signal interactions, show considerable dependence on the modulation formats applied.

Furthermore, the minimum required number of steps per span in the MC-NLC algorithm has been investigated in terms of the AIRs in the compensated communication systems and were compared to the cases where the SNRs are optimised. It is observed that in the AIR optimisation the minimum required number of steps per span at different back-propagated bandwidths strongly depends on the modulation format applied, in contrast to the scheme of SNR optimisation where for a given back-propagated bandwidth the number of steps per span required is the same for all modulation formats. Our investigation shows that, for a given AIR, there exists a potential trade-off between the modulation format and back-propagated bandwidth and a compromise may be achieved according to practical complexity limitations.

This paper gives an insight on the optimisation approaches of digital multi-channel NLC, and the selection of modulation formats and back-propagated bandwidths to optimise practical optical fibre communication systems.

ACKNOWLEDGEMENTS

UK EPSRC project UNLOC (EP/J017582/1) and EU Marie Skłodowska-Curie project COIN (676448/H2020-MSCA-ITN-2015).

REFERENCES

[1] M. Secondini, E. Forestieri, and G. Prati, "Achievable information rate in nonlinear WDM fiber-optic systems with arbitrary modulation formats and dispersion maps," J. Lightwave Technol., vol. 31, 2013, pp. 3839-3852.

[2] L. Szczecinski and A. Alvarado, Bit-interleaved coded modulation: fundamentals, analysis and design. New Jersey: John Wiley & Sons, Inc., 2015.

[3] D. Semrau, T. Xu, N. A. Shevchenko, M. Paskov, A. Alvarado, R. I. Killey, and P. Bayvel, "Achievable information rates estimates in optically amplified transmission systems using nonlinearity compensation and probabilistic shaping," Opt. Lett., vol. 42, 2017, pp. 121-124.

[4] S. Dimitrov and H. Haas, "Information rate of OFDM-based optical wireless communication systems with nonlinear distortion," J. Lightwave Technol., vol. 31, 2013, pp. 918-929.

[5] B. Mukherjee, "WDM optical communication networks: progress and challenges," IEEE J. Sel. Areas Commun., vol. 18, 2000, pp. 1810-1824.

[6] G. P. Agrawal, Fiber-optic communication systems, 4th ed. New Jersey: John Wiley & Sons, Inc., 2010.

[7] T. Xu, G. Jacobsen, S. Popov, J. Li, K. Wang, A. T. Friberg, "Digital compensation of chromatic dispersion in 112-Gbit/s PDM-QPSK system," Digests Asia Commun. Photon. Conf., p. TuE2, 2009.

[8] P. J. Winzer and R.-J. Essiambre, "Advanced optical modulation formats," Proc. IEEE, vol. 94, 2006, pp. 952-985.

[9] X. Liu, S. Chandrasekhar, and P. J. Winzer, "Digital signal processing techniques enabling multi-Tb/s superchannel transmission: an overview of recent advances in DSP-enabled superchannels," IEEE Signal Process. Mag., vol. 31, 2014, pp. 16-24.

[10] G. Jacobsen, T. Xu, S. Popov, J. Li, A. T. Friberg, and Y. Zhang, "Phase noise influence in coherent optical OFDM systems with RF pilot tone: digital IFFT multiplexing and FFT demodulation," J. Opt. Commun., vol. 33, 2012, pp. 217-226.

[11] N. A. Olsson, "Lightwave systems with optical amplifiers," J. Lightwave Technol., vol. 7, 1989, pp. 1071-1082.

[12] M. S. Faruk and S. J. Savory, "Digital signal processing for coherent transceivers employing multilevel formats," J. Lightwave Technol., vol. 35, 2017, pp. 1125-1141.

[13] E. Ip, A. P. T. Lau, D. J. F. Barros, and J. M. Kahn, "Coherent detection in optical fiber systems," Opt. Express, vol. 16, 2008, 753-791.

[14] T. Xu, G. Jacobsen, S. Popov, J. Li, S. Sergeyev, A. T. Friberg, Y. Zhang, "Analytical BER performance in differential n-PSK coherent transmission system influenced by equalization enhanced phase noise," Opt. Commun., vol. 334, 2015, pp. 222-227.

[15] R.-J. Essiambre and R. W. Tkach, "Capacity trends and limits of optical communication networks," Proc. IEEE, vol. 100, 2012, pp. 1035-1055.

[16] P. P. Mitra and J. B. Stark, "Nonlinear limits to the information capacity of optical fibre communications," Nat., vol. 411, 2001, pp. 1027-1030.

[17] G. Keiser, Optical fiber communications. New Jersey: John Wiley & Sons, Inc., 2003.

[18] T. Xu, J. Li, G. Jacobsen, S. Popov, A. Djupsjöbacka, R. Schatz, Y. Zhang, and P. Bayvel, "Field trial over 820km installed SSMF and its potential Terabit/s superchannel application with up to 57.5-Gbaud DP-QPSK transmission," Opt. Commun., vol. 353, 2015, pp. 133-138.

[19] G. P. Agrawal, Nonlinear fibre optics, 5th ed. Massachusetts: Academic Press, 2013.

[20] A. D. Ellis, M. E. McCarthy, M. A. Z. Al Khateeb, M. Sorokina, and N. J. Doran, "Performance limits in optical communications due to fiber nonlinearity," Adv. Opt. Photon., vol. 9, 2017, pp. 429-503.

[21] E. M. Ip and J. M. Kahn, "Compensation of dispersion and nonlinear impairments using digital backpropagation," J. Lightwave Technol., vol. 26, 2008, pp. 3416-3425.

[22] J. C. Cartledge, F. P. Guiomar, F. R. Kschischang, G. Liga, and M. P. Yankov, "Digital signal processing for fiber nonlinearities," Opt. Express, vol. 25, 2017, pp. 1916-1936.

[23] E. Temprana, E. Myslivets, B.P.-P Kuo, L. Liu, V. Ataie, N. Alic, and S. Radic, "Overcoming Kerr-induced capacity limit in optical fiber transmission," Sci., vol. 348, 2015, pp. 1445-1448.

[24] F. P. Guiomar, J. D. Reis, A. L. Teixeira, and A. N. Pinto, "Mitigation of intra-channel nonlinearities using a frequency-domain Volterra series equalizer," Opt. Express, vol. 20, 2012, pp. 1360-1369.

[25] I. D. Phillips, M. Tan, M. F. C. Stephens, M. E. McCarthy, E. Giacoumidis, S. Sygletos, P. Rosa, S. Fabbri, S. T. Le, T. Kanesan, S. K. Turitsyn, N. J. Doran, P. Harper, and A. D. Ellis, "Exceeding the nonlinear-Shannon limit using Raman laser based amplification and optical phase conjugation," Digests Opt. Fiber Commun. Conf., p. M3C.1, 2014.

[26] M. I. Yousefi and F. R. Kschischang, "Information transmission using the nonlinear Fourier transform, Part I-III: Numerical methods," IEEE Trans. Inf. Theory, vol. 60, 2014, pp. 4329-4345.

[27] X. Liu, A. R. Chraplyvy, P. J. Winzer, R. W. Tkach, and S. Chandrasekhar, "Phase-conjugated twin waves for communication beyond the Kerr nonlinearity limit," Nat. Photon., vol. 7, 2013, pp. 560-568.

[28] D. Rafique, "Fiber nonlinearity compensation: commercial applications and complexity analysis," J. Lightwave Technol., vol. 34, 2016, pp. 544-553.

[29] A. Napoli, Z. Maalej, V. A. J. M. Sleiffer, M. Kuschnerov, D. Rafique, E. Timmers, B. Spinnler, T. Rahman, L. D. Coelho, and N. Hanik, "Reduced complexity digital back-propagation methods for optical communication systems," J. Lightwave Technol. vol. 32, 2014, pp. 1351-1362.

[30] L. B. Du and A. J. Lowery, "Improved single channel backpropagation for intra-channel fiber nonlinearity compensation in long-haul optical communication systems," Opt. Express, vol. 18, 2010, pp. 17075-17088.

[31] B. Karanov, T. Xu, N. A. Shevchenko, D. Lavery, R. I. Killey, and P. Bayvel, "Span length and information rate optimisation in optical transmission systems using single-channel digital backpropagation,", Opt. Express, vol. 25, 2017, in press.

[32] N. K. Fontaine, X. Liu, S. Chandrasekhar, R. Ryf, S. Randel, P. Winzer, R. Delbue, P. Pupalakis, and A. Sureka, "Fiber nonlinearity compensation by digital backpropagation of an entire 1.2 Tb/s superchannel using a full-field spectrally-sliced receiver," Digests Opt. Fiber Commun. Conf., p. Mo.3.D.5, 2013.

[33] G. Liga, T. Xu, A. Alvarado, R. I. Killey, and P. Bayvel, "On the performance of multichannel digital backpropagation in high-capacity long-haul optical transmission," Opt. Express, vol. 22, 2014, pp. 30053-30062.

[34] R. Maher, T. Xu, L. Galdino, M. Sato, A. Alvarado, K. Shi, S. J. Savory, B. C. Thomsen, R. I. Killey, and P. Bayvel, "Spectrally shaped DP-16QAM super-channel transmission with multi-channel digital back-propagation," Sci. Rep., vol. 5, 2015, pp. 8214.

[35] T. Xu, G. Liga, D. Lavery, B. C. Thomsen, S. J. Savory, R. I. Killey, and P. Bayvel, "Equalization enhanced phase noise in Nyquist-spaced superchannel transmission systems using multi-channel digital back-propagation," Sci. Rep., vol. 5, 2015, pp. 13990.

[36] C. Xia, X. Liu, S. Chandrasekhar, N. K. Fontaine, L. Zhu, and G. Li, "Multi-channel nonlinearity compensation of PDM-QPSK signals in dispersion-managed transmission using dispersion-folded digital backward propagation," Opt. Express, vol. 22, 2014, pp. 5859-5866.

[37] A. Amari, O. A. Dobre, R. Venkatesan, O. S. S. Kumar, P. Ciblat, and Y. Jaouën, "A survey on fiber nonlinearity compensation for 400 Gbps and beyond optical communication systems," IEEE Commun. Surv. Tutorials, 2017, in press.

[38] T. Xu, N. A. Shevchenko, D. Lavery, D. Semrau, G. Liga, A. Alvarado, R. I. Killey, and P. Bayvel, "Modulation format dependence of digital nonlinearity compensation performance in optical fibre communication systems," Opt. Express, vol. 25, 2017, pp. 3311-3326.

[39] R. Dar and P. J. Winzer, "On the limits of digital back-propagation in fully loaded WDM systems, IEEE Photon. Technol. Lett., vol. 28, 2016, pp. 1253-1256.

[40] T. Xu, B. Karanov, N. A. Shevchenko, D. Lavery, G. Liga, R. I. Killey, and P. Bayvel, "Digital nonlinearity compensation in high-capacity optical communication systems considering signal spectral broadening effect," Sci. Rep., 2017, in press.

[41] G. Bosco, A. Carena, V. Curri, R. Gaudino, P. Poggiolini, and S. Benedetto, "Suppression of spurious tones induced by the split-step method in fiber systems simulation," IEEE Photon. Technol. Lett., vol. 12, 2000, pp. 489-491.

[42] T. Xu, G. Jacobsen, S. Popov, J. Li, E. Vanin, K. Wang, A. T. Friberg, and Y. Zhang, "Chromatic dispersion compensation in coherent transmission system using digital filters," Opt. Express, vol. 18, 2010, pp. 16243-16257.

[43] T. Xu, G. Jacobsen, S. Popov, M. Forzati, J. Mårtensson, M. Mussolin, J. Li, Y. Zhang, and A. T. Friberg, "Frequency-domain chromatic dispersion equalization using overlap-add methods in coherent optical system," J. Opt. Commun. vol. 32, 2011, pp. 131-135.

[44] P. Serena, A. Bononi, and N. Rossi, "The impact of the modulation dependent nonlinear interference missed by the Gaussian noise model," Digests European Conf. Opt. Commun., p. Mo.4.3.1, 2014.

[45] R. Dar, M. Feder, A. Mecozzi, and M. Shtaif, "Accumulation of nonlinear interference noise in fiber-optic systems," Opt. Express, vol. 22, 2014, pp. 14199-14211.

[46] A. Carena, G. Bosco, V. Curri, Y. Jiang, P. Poggiolini, and F. Forghieri, "EGN model of non-linear fiber propagation," Opt. Express, vol. 22, 2014, pp. 16335-16362.